\documentstyle[mnras_cite,epsfig,eqnarray]{mn2e}

\def\gsim{~\rlap{$>$}{\lower 1.0ex\hbox{$\sim$}}}
\def\lsim{~\rlap{$<$}{\lower 1.0ex\hbox{$\sim$}}} \def\G{{\rm G}}
\def\G{{\rm G}}
\def\c{{\rm c}}
\def\d{{\rm d}}

\newfont{\boldit}{cmbxti10 scaled \magstep 2}

\title{Maximum Spin of Black Holes Driving Jets}
\author[Andrew J. Benson \& Arif Babul]{Andrew J. Benson$^1$ and Arif Babul$^2$\\
$^1$Mail Code 130-33, California Institute of Technology, Pasadena, CA~91125, U.S.A. (e-mail: {\tt abenson@caltech.edu})\\
$^2$Department of Physics \& Astronomy, University of Victoria, Elliott Building, 3800 Finnerty Road, Victoria, BC V8P 1A1, Canada}

\begin{document}

\maketitle

\begin{abstract}
Unbounded outflows  in the form of highly collimated jets and broad winds appear to be a ubiquitous feature of accreting black hole systems.  The most powerful jets are thought to derive a significant fraction, if not the majority, of their power from the rotational energy of the black hole. Whatever the precise mechanism that causes them, these jets must therefore exert a braking torque on the black hole. Consequently, we expect jet production to play a significant role in limiting the maximum spin attainable by accreting black holes.
We calculate the spin-up function---the rate of change of black hole spin normalized to the black hole mass and accretion rate---for an accreting black hole, accounting for this braking torque. We assume that the accretion flow onto a Kerr black hole is advection-dominated (ADAF) and construct easy-to-use analytic fits to describe the global structure of such flows based on the numerical solutions of Popham \& Gammie (1998).  We find that the predicted black hole spin-up function depends only on the black hole spin and dimensionless parameters describing the  accretion flow.  Using recent relativistic magnetohydrodynamical (MHD) numerical simulation results to calibrate the efficiency of angular momentum transfer in the flow, we find that an ADAF flow will spin a black hole up (or down) to an equilibrium value of about $96\%$ of the maximal spin value in the absence of jets. Combining our ADAF system with a simple model for jet power, we demonstrate that an equilibrium is reached at approximately $93\%$ of the maximal spin value, as found in the numerical  simulation studies of the spin-up of accreting black holes, at which point the spin-up of the hole by accreted material is balanced by the braking torque arising from jet production.  The existence of equilibrium spin means that optically-dim AGNs that have grown via accretion from an advection-dominated flow will not be maximally rotating.   It also offers a possible explanation for the tight correlation observed by \scite{allen06} between the Bondi accretion rate and jet power in nine, nearby, X-ray luminous giant elliptical galaxies.   We suggest that the black holes in these galaxies must all be rotating close to their equilibrium value.   Our model also yields a relationship between jet efficiency and black hole spin that is in surprisingly good agreement with that seen in the simulation studies, indicating that our simple model is a useful and convenient description of ADAF inflow -- jet outflow about a spinning black hole for incorporation in semi-analytic modeling as well as cosmological numerical simulation studies focusing on the formation and evolution of galaxies, groups and clusters of galaxies.        
\end{abstract}


\section{Introduction}

Astrophysical black holes are characterized by just two properties, their mass and angular momentum, since they will typically have zero charge \cite{BZ77}. The angular momentum, $J$, of a Kerr black hole of mass $M_\bullet$, can be defined in terms of a dimensionless spin $j=Jc/GM_\bullet^2$ which must lie in the range $-1< j < 1$. There is mounting evidence that astrophysical black holes are rotating and that the corresponding spin is a critical variable in establishing the observed properties of the accreting black hole systems (see discussion in \pcite{mg04,gsm04}) as well as the magnitude and the impact that the outflows from these systems will have on their surroundings---see \scite{nemmen07} and references therein. The energy associated with this spin can be up to 30\% of the gravitational mass in an extreme Kerr black hole \cite{mg04} and is, in principle, available for extraction \cite{penrose69} and so is a likely power source for phenomena such as jets from active galactic nuclei, microquasars and gamma ray bursts. Understanding how black holes acquire their spins and, in the cosmological context, what the distribution of spins is, is therefore of fundamental importance. 

Black holes can gain angular momentum either via mergers with other black holes or from the material that they accrete.  In this paper, we restrict our discussion to accretion. \scite{bardeen70} was the first to consider the issue.   Assuming that the accretion proceeded via a cold disk and that the angular momentum of the accreting material was always aligned with the spin of the black hole, he noted that an accreting black hole would spin up to its maximal rate in a finite time (i.e. after accreting around 1.5 times the initial black hole mass).  \scite{thorne74} recognized that thin disks radiate and as a result, the black hole spin would be limited to  $j\approx 0.998$ because the black hole would preferentially swallow negative angular momentum photons emitted by the accretion flow.  A number of studies since then have suggested that this limit may not hold (or that the limiting spin may be higher) if accretion takes place via a geometrically thick flow.  \scite{popgam98}, who numerically solved the equations describing the steady state radial structure of hot, geometrically thick, relativistic, optically thin, advection-dominated accretion flow, found that the equilibrium spin can be as high as $0.99997$ or as low as $0.7$, depending on the efficiency of angular momentum transport in the flow; they did not, however, take into account radiation swallowing.  In the case of pure advection flows where none of the energy dissipated in the flow is radiated away, this is not an issue.   Finally, as noted by J.~M.~Bardeen (related by \pcite{thorne74}), the effects of magnetic fields could potentially limit the spin to a significantly lower value by exerting torques on the accreting material. 

The potential role of magnetic fields has received increasing attention in the past decade, beginning with the works of \scite{krolik99} and \scite{gammie99}, who demonstrated that the magnetic fields expected on the basis of flux freezing should have order unity effects on the accretion flow and therefore, on the amount of energy released in that flow. The effects of magnetic fields have been studied further with the general consensus that magnetic stresses can significantly alter the spin-up of the black hole \cite{li00,li02,agolkrolik00}. In particular, \scite{agolkrolik00} used the second law of black hole thermodynamics to place constraints on the maximum accretion efficiency for a given black hole spin. They found a maximum efficiency of 0.36, which would be attained if the equilibrium spin was 0.94.

In recent years, the use of sophisticated general relativistic, magnetohydrodynamic (MHD) numerical simulations (e.g., \pcite{koide00,koide03,deV03,mg04,gsm04,komissarov_observations_2005,vhkh05,hawley06,komissarov_magnetic_2007,punsly07,beckwith08,mckinney09}) have yielded considerable insights into the role of magnetic fields in  accretion flows, and especially the efficacy of magnetic stresses at extracting rotational energy from a black hole-accretion flow system.  In the specific simulations presented by \scite{mg04} and \scite{gsm04},  stresses associated with magnetic fields in the accretion flow resulted in the black hole evolving towards an equilibrium spin (i.e. $\d j/\d t=0$) of $j\sim 0.9$ (specifically equilibrium occurs somewhere between their $j=0.90$ and $j=0.94$ simulations). \scite{gsm04} note that this equilibrium value is relevant only to geometrically thick accretion flows.  \scite{khh05} too reach a similar conclusion (i.e. that there is an equilibrium spin at approximately $j=0.9$) by analyzing the MHD simulations of \scite{deV03}. A more in-depth review of these studies can be found in \scite{krolik05}. To summarize, current MHD simulations of spinning black holes accreting from a geometrically thick accretion flow predict an equilibrium spin somewhere in the range $j=0.9$--$0.94$,  the exact value being uncertain due to numerical limitations and the small number of simulations available. At higher values of $j$, the black hole is spun-down.

A key feature of all of these MHD simulations is the emergence of unbounded outflows.   These outflows can occur in the form of a highly collimated, Poynting-flux dominated component generated both by the magnetized, rapidly rotating accretion flow as well as the rotation of the black hole \cite{vhkh05,punsly07,mckinney09}, and as a broad, mildly relativistic, typically matter-dominated, component that originates in the accretion flow\footnote{McKinney and collaborators refer to this component as the ``disk wind'' (c.f. \pcite{mckinney06,mckinney09})}.   In the present paper, we will use the generic label ``jet'' to refer to the outflows, regardless of whether they be broad or narrow, matter or Poynting-flux dominated, and instead we will distinguish between outflow engendered by the magnetic fields anchored in the rotating flow (hereafter referred to as the ``disk'' jet) and that due to field lines anchored on the black hole event horizon (the ``black hole'' jet). The black hole jet is primarily electromagnetic and the disk jet is hydromagnetic, where the energy (and the angular momentum) is carried by an electromagnetic flux as well as a kinetic flux of matter.

While a detailed, universally accepted, description of the mechanisms underlying the origin and power of these jets remains elusive, the combination of the results from simulations with different parameters and analysis of simplified analytic models, suggests the following tentative picture that we use as a basis for our model for the jets (see Appendix ~\ref{app:jet}):  

As a starting point, we are interested in black holes situated at centres of galaxies and clusters of galaxies. In such systems, the gas accreting onto the black hole will have 
originally flowed in from kiloparsecs-scale via a cooling flow and a 
quasi-spherical accretion flow. During the latter stage, the 
magnetothermal instability is expected to organize the existing weak 
magnetic fields into primarily radial fields (e.g.~\pcite{sharma08}). We will therefore assume that there is  a weak, large-scale poloidal component present in the accretion flow onto the black hole.    Although the detailed geometric structure of the flow as it converges onto the black hole will depend on whether the gas is able to radiate away its thermal energy, generically the flow will be rotating and axisymmetric in character.   In the present paper, we will focus primarily on geometrically thick, advection-dominated accretion flows (ADAFs).

Several MHD processes are expected to arise in the flow:  the shearing of the poloidal field within the rotating fluid will give rise to a toroidal field and the flow will be subject to MHD turbulence.  These features have been confirmed by the various simulations.   The magnetic fields anchored in the rotating flow will give rise to a combination of outflow of material from the disk and its surroundings, and an electromagnetic flux, with the former dominating if the black hole is spinning relatively slowly.  As recognized by (\pcite{meier99}; see also \pcite{meier01}), if the black hole is spinning, the frame-dragging of the inflowing gas within the ergosphere will enhance the magnitude of the outflow.  In fact,  according to \scite{punsly07}, not only do the numerical simulations by \scite{hawley06} confirm that the power of the hydromagnetic disk jet indeed grows with black hole spin, but that the rise in power is very steep at high black hole spin rates and that this steep increase is primarily due to growth in the power of the electromagnetic component of the disk jet.

Finally, when the accretion flow reaches the black hole event horizon, the gas is expected to drain off the field lines.  In the case of a non-rotating black hole, the  magnetic pressure will cause the field lines, which for all intents and purposes can be thought of as being anchored on the black hole's event horizon, to establish a nearly radial configuration in the polar regions similar to the split-monopole structure first described by \scite{BZ77}.   In the event that the black hole is spinning,  the winding of the magnetic field lines in the ergosphere will drive helical twists along the magnetic tower that manifests as highly collimated, Poynting-flux dominated jets.  

In the picture outlined above, the combined disk+black hole jet therefore draws its energy from the gravitational energy released by the accretion flow as well as  the rotational energy of the black hole itself.  Consequently, the jet power does not vanish in the case of a Schwarzchild black hole.   This is consistent with simulation results in that studies exploring the relationship between the outflow structure and power, and the spin of the black hole (e.g. \pcite{mg04,vhkh05,hawley06}) all find that disk winds persist even when the black hole is not rotating.   However, in cases where the magnetic field lines are  appropriately orientated\footnote{Recent simulation studies suggest that an ordered polodial component is optimal (c.f., \pcite{beckwith08,mckinney09}).},  the presence of a rapidly spinning black hole can greatly enhance the outflow power.  This link between black hole spin and outflow power has long been indicated from theoretical considerations  (c.f., \pcite{BZ77,punsly90,meier99,meier01}).

Given that the jets gain a  fraction of their power by tapping the rotational energy of the black hole, we can estimate the braking torque exerted on the black hole and consequently, the maximum spin attainable by an accreting black hole driving powerful jets.  Utilizing the understanding of jet production developed by these simulations, we construct a simple, analytical model for the maximum spin attainable through accretion. To do this, we assume an advection-dominated accretion flow (ADAF) onto a spinning black hole whose global structure resembles the numerical solutions of  \scite{popgam98}, and we examine how the accretion-driven spin-up of a black hole is modified when jets are produced as a consequence of that accretion.  Using a simple model for the jet power, where a fraction of the jet power is extracted from the rotational energy of the black hole, we derive a modified spin-up function and thereby predict the equilibrium value of the black hole spin.  In the process, we also determine the dependence of jet efficiency as a function of black hole spin.

\section{Spin-up of a Black Hole}

We wish to compute the rate of spin-up for a black hole accreting from a geometrically thick, advection dominated accretion flow (ADAF) that also engenders powerful jets. We will begin by examining the classical calculation of the spin up of a black hole accreting from a thin disk and with no jet production. We will then proceed to modify this calculation to describe accretion from an ADAF and finally, quantify how jet production alters this spin-up.

\subsection{Spin-up by a thin accretion disk}

As stated above, we quantify the angular momentum of a Kerr black hole by the dimensionless parameter $j$ $(\equiv Jc/GM_\bullet^2)$. We follow the notation of \scite{shapiro05} and define a dimensionless spin-up function $s(j)$ by
\begin{equation}
s(j) = {\d j\over \d t} {M_\bullet \over \dot{M}_{\bullet,0}},
\label{eq:spinup}
\end{equation}
where $\dot{M}_{\bullet,0}$ is the rate of rest mass accretion. For a standard, relativistic, Keplerian thin-disk accretion flow with no magnetic fields, we will denote the expected spin-up function due to accretion by $s_0(j)$. \scite{shapiro05} gives
\begin{equation}
s_0(j) = {\mathcal L}_{\rm ISCO}-2 j E_{\rm ISCO},
\end{equation}
where
\begin{eqnarray}
{\mathcal L}_{\rm ISCO}(j) & = & {\sqrt{r_{\rm ISCO}} (r_{\rm ISCO}^2-2j\sqrt{r_{\rm ISCO}}+j^2) \over r_{\rm ISCO} \sqrt{r_{\rm ISCO}^2-3r_{\rm ISCO}+2j\sqrt{r_{\rm ISCO}}}} \\
E_{\rm ISCO}(j) & = & {r_{\rm ISCO}^2-2r_{\rm ISCO}+j\sqrt{r_{\rm ISCO}} \over r_{\rm ISCO} \sqrt{r_{\rm ISCO}^2-3r_{\rm ISCO}+2j\sqrt{r_{\rm ISCO}}}},
\end{eqnarray}
are the (dimensionless) specific angular momentum and specific energy of the innermost stable circular orbit (ISCO) of the black hole respectively.   This relationship implicitly assumes that (a) the inner edge of the disk coincides approximately with the radius of the innermost stable circular orbit, inside of which the centrifugal force is unable to balance gravity and the gas begins to free-fall inward, and (b) that the torque at this inner boundary is negligible since the gas in the plunging region will quickly accelerate to supersonic speeds and lose causal contact with the material upstream.   This ``no-torque'' boundary condition has been a subject of some debate but both \scite{afpac03} and \scite{li03} have independently shown that in the case of thin disks, the assumption is reasonable. Moreover, recent simulations \cite{shafee08} confirm that the torque is negligible (at least for $j\approx 0$).
The radius of the ISCO orbit (in units of the gravitational radius $\G M_\bullet/\c^2$) is
\begin{equation}
r_{\rm ISCO}=  3+A_2(j)-\sqrt{[3-A_1(j)][3+A_1(j)+2A_2(j)]}
\label{eq:isco}
\end{equation}
where
\begin{equationarray}{lcl}
A_1(j) & = & 1+[(1-j^2)^{1/3}][(1+j)^{1/3}+(1-j)^{1/3}],\cr
A_2(j) & = & \sqrt{3j^2+A_1(j)^2}.
\end{equationarray}

\begin{figure*} 
\centering 
\includegraphics[width=80mm]{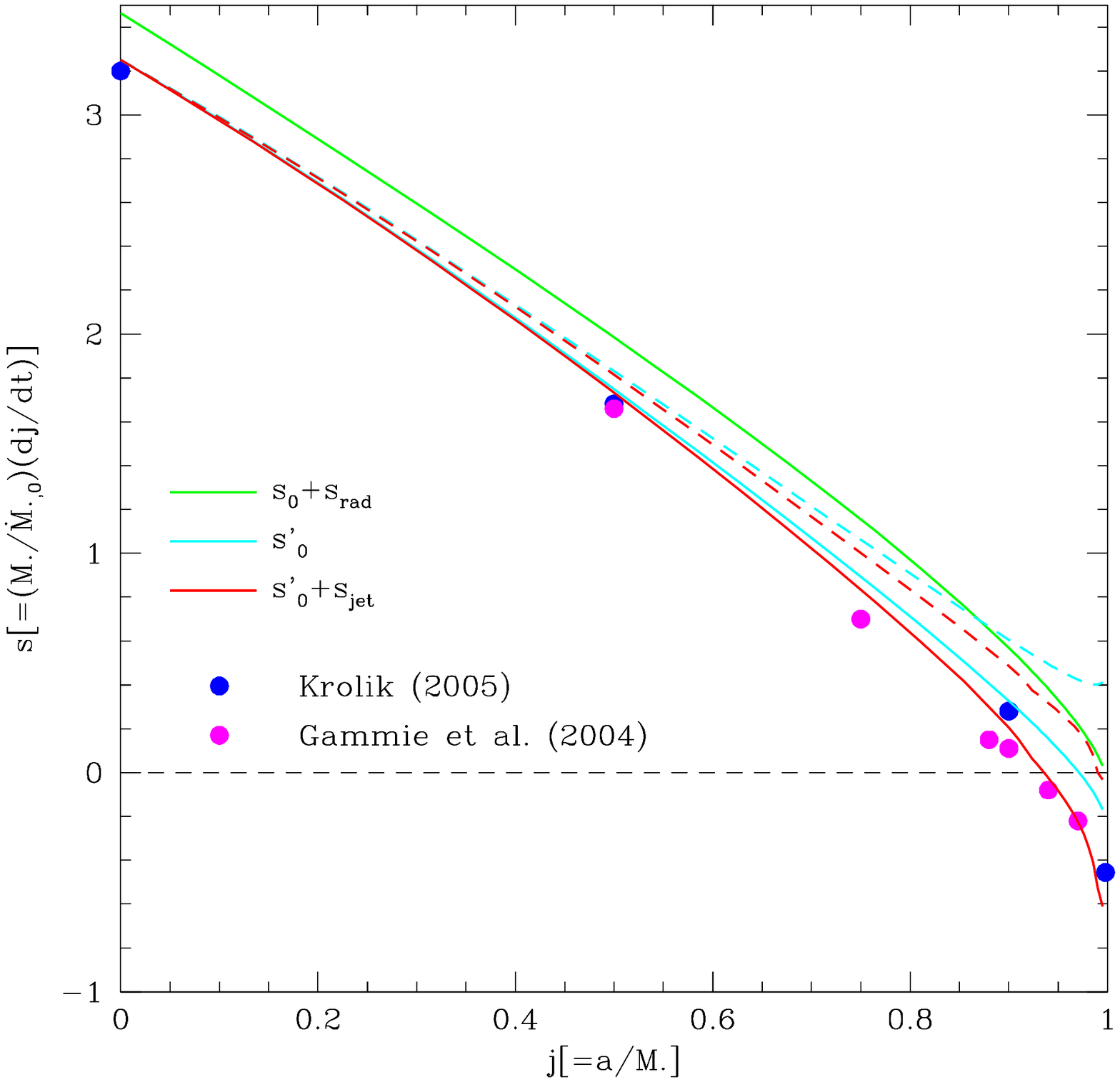}%
\hspace{1cm}%
\includegraphics[width=80mm]{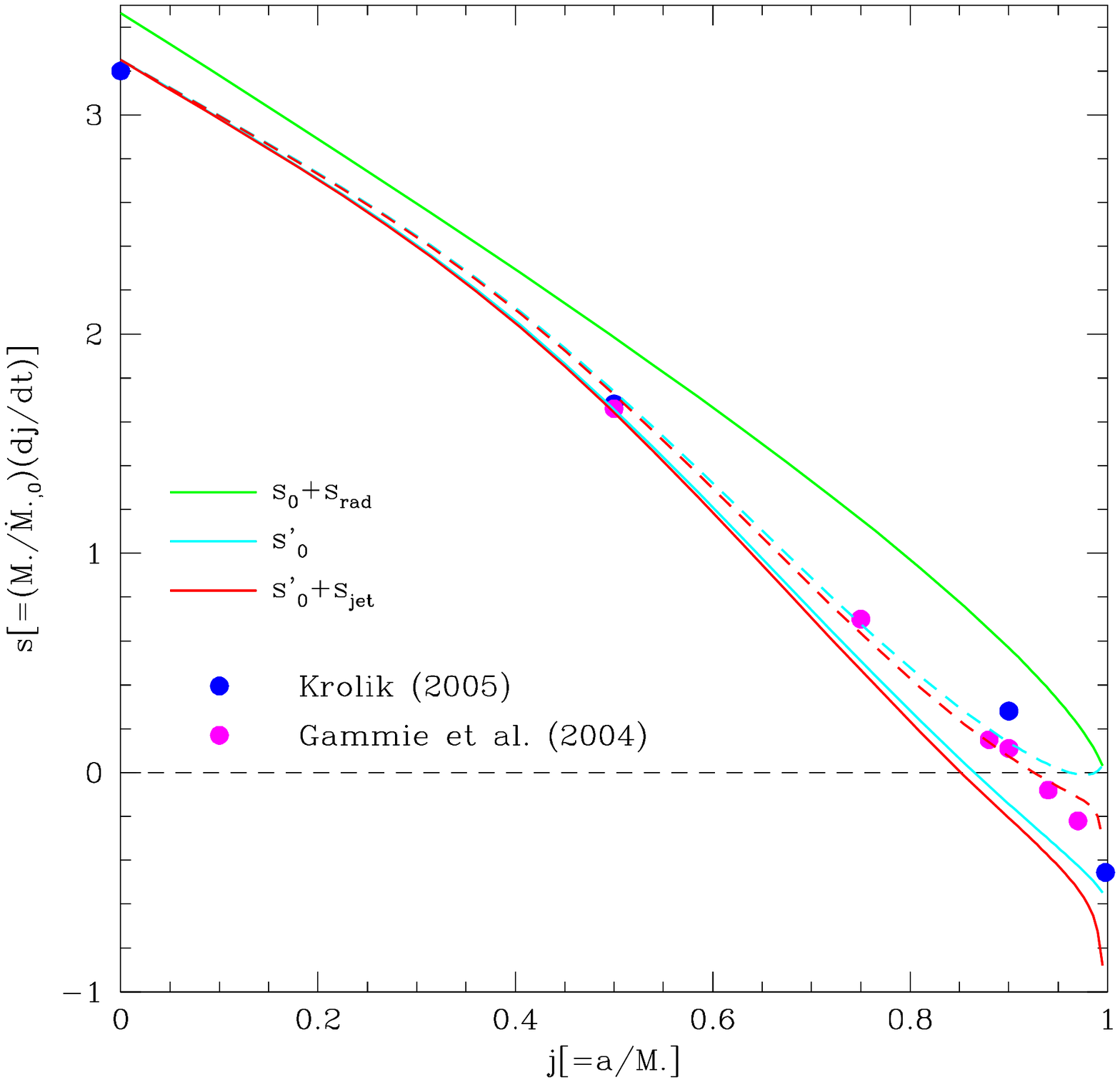} 
\caption{Spin-up parameter of a black hole as a function of spin, $j$, for the case $f=1$ and $\gamma=1.44$. Lines show results from the present work. The green line corresponds to spin-up by a thin disk (i.e. $s_0$) with the radiation swallowing term, $s_{\rm rad}$, included (but no braking term due to jet power). The blue line corresponds to spin-up by an ADAF (i.e. $s_0^\prime$) without the radiation swallowing term or jet braking term included, while the red line includes the jet braking term. We do not factor in the $s_{\rm rad}$ term for the ADAF cases since these flows are radiatively inefficient and $s_{\rm rad}$ is small for all $j$ (i.e. removing it would make no noticeable difference to the lines in this figure). The left plot uses $\alpha(j;E=1)$ and the right plot is based on $\alpha(j;E=E_{\rm ISCO})$.  The spin-up function for an ideal ADAF ($E=1$) is shown as a solid curve while the results for $E=E_{\rm ISCO}$ appear as dashed curves.   Also shown are the simulation results from \protect\scite{krolik05} and \protect\scite{gsm04} (blue and magenta points respectively). The horizontal line indicates the equilibrium state where $s=0$.}
\label{fig:spineq}
\end{figure*}

The net spin-up rate in the case of thin disk accretion is given by $s_0(j) + s_{\rm rad}$, where $s_{\rm rad}$ is the spin-down rate due to radiation swallowing as given by \scite{thorne74}\footnote{Note that the quantity $\d a_\ast/\d\ln M$ plotted in Figure~6 of \protect\scite{thorne74} is defined using the actual change in mass of the black hole rather than mass flow rate in the accretion flow and therefore differs from our quantity $s$ by a factor of $E_{\rm ISCO}$.}.  As illustrated in Figure~\ref{fig:spineq}, the net spin-up function for the standard thin-disk (green curve) is positive for all $j<0.998$, and therefore lets the black hole spin up to $j\approx 0.998$ in finite time as noted by \scite{shapiro05}\footnote{Shapiro  did not include the radiation breaking term and so finds that the black hole spins up to $j=1$. We note that  the radiation braking term is small for all $j$ and discounting it would make no noticeable difference to any of the curves in this Figure.}.

\subsection{Spin-up by an ADAF}

Thin disks are expected to be features of radiatively efficient accretion flows and therefore, typically associated with optically luminous active galactic nuclei.   The black hole systems that are of particular interest to us, i.e., supermassive black holes that reside at the centres of massive elliptical galaxies, are typically optically-dim.  In the inner regions, these systems are thought to be accreting in an optically thin, advection-dominated mode, in which the density of the accreting gas is sufficiently low that the gas cannot radiate efficiently;  the energy dissipated during the accretion remains in the flow in the form of thermal energy and is eventually carried across the horizon \cite{emn97,narayan98,nm08}.

Whether the accretion flow is geometrically thin and radiatively efficient, or geometrically thick, optically thin, low luminosity and advection-dominated, or for that matter, geometrically thick, optically thick, luminous but yet advection-dominated, depends largely on the mass accretion rate.  Estimates of mass accretion rates onto the supermassive black holes in massive elliptical galaxies (c.f., \pcite{allen06}) yield very low values and 
at low mass accretion rates, the flow is expected to be an optically thin, low-luminosity ADAF.\footnote{Since we are specifically interested in the properties of optically dim AGNs in massive elliptical galaxies, we will not consider the luminous, advection-dominated flows any further.  Henceforth, we will use the label ``ADAF'' as a short-form for ``optically-thin, low-luminosity, advection-dominated flows". Such flows are also sometimes referred to as radiatively inefficient accretion flows.}

The physical structure of an ADAF is very different from that of a thin disk.   Consequently, the thin-disk expression for $s_0$ is unlikely to be correct for an ADAF.  A priori, one would expect both the specific angular momentum and the specific energy of the accreted material to differ from the thin-disk expectations. We therefore write the spin-up rate for ADAFs as:
\begin{equation}
 s_0^\prime = {\mathcal L}_{\rm ADAF}(j,\alpha) - 2 j E_{\rm ADAF},
\end{equation}
where ${\mathcal L}_{\rm ADAF}$ and $E_{\rm ADAF}$ are normally evaluated at the horizon.

In the case of an ideal, unmagnetized ADAF, in which the gas starts out cold at large radii and all dissipated energy is advected along with the flow, the (dimensionless) specific energy of the accreted material is  $E_{\rm ADAF} \sim 1$ \cite{popgam98}, while the specific angular momentum of the accreted gas is typically sub-Keplerian \cite{narayan97,popgam98,manmoto00}.   Recent simulations, which include the effects of magnetic fields and allow for the exploration of less constrained flow structures, confirm the expectation for the angular momentum, that it is typically  less than the thin-disk result, but suggest that the energy of the accreted material is in fact close to the thin-disk value (see Figure~5 of \pcite{mg04} and also \pcite{vhkh05}).  In this paper, we will show results for  both values for the specific energy of the accreting material, the thin-disk value as indicated by the simulations as well as the ideal ADAF value.

To quantify the accreted angular momentum (and, as described in the next section, to estimate jet power), we need a model for the ADAF accretion flow.   To this end, we make use of the results of \scite{popgam98}, who numerically solved the equations describing the radial structure of steady state, optically thin, advection-dominated accretion flow in a Kerr metric (see also, \pcite{gampop98}).    Though more cumbersome to use than the simple, easy-to-manipulate, self-similar solutions of \scite{ny95,narayan98,bu2008}, they offer one key advantage over the latter: an improved treatment of the flow in the innermost regions of the ADAF where the self-similar results break down. The resulting ADAF model includes the effects of magnetic fields to the extent that the flow is treated as a mixture of a frozen-in, isotropically tangled, turbulent magnetic fields and ionized plasma.  It does not, however, account for large-scale magnetic fields or MHD effects, like the amplification of magnetic fields by dynamo-like processes and MHD winds.  On the other hand, \scite{popgam98} explicitly illustrate how the flow structure depends on the black hole spin $j$,  the adiabatic index $\gamma$ quantifying the strength of the turbulent magnetic fields in the flow, the viscosity parameter $\alpha$ governing the transport of angular momentum in the flow, and the advected fraction of the dissipated energy $f$.  

For the purposes at hand, we adopt their $f=1$ (pure advection flow) results and to facilitate the use of these solutions in this (and future) work, we have derived analytic fits to key ADAF structural quantities, from which additional properties of the flow can be derived.   These simple fitting functions are listed in Appendix~\ref{app:adaf}.  For simplicity's sake, we shall assume that the rate of angular momentum accretion by the black hole per unit rest mass accreted, ${\mathcal L}_{\rm ADAF}$, is equal to the value of the specific angular momentum of the accreting fluid near the horizon.   This is tantamount to assuming a ``no-torque'' boundary condition at the horizon.  Strictly speaking, this condition does not apply to geometrically thick flows but in the model under consideration, the rate of angular momentum transport is very small near the horizon and the two quantities agree to within $\sim$20\%.

In Popham and Gammie's treatment, the viscous stress is assumed to result from the combined effect of correlated, large-scale, time-averaged Reynolds and Maxwell stresses  that they associate with MHD turbulence.   They model this stress using a modified form of the   $\alpha$-viscosity prescription designed to ensure causal behaviour and their $\alpha$ parameter is defined in terms of the total pressure, which includes the contribution of the turbulent magnetic field pressure.    \scite{popgam98}  assume that $\alpha$, as well as all other parameters mentioned above, are constant over the radial extent of the flow.   MHD simulation studies of magnetized accretion flows suggest that this approach does not do justice to the rich dynamics involved.  For one,  MHD simulations of accretion flows around black holes have long indicated that the effective value of $\alpha$ is a strong function of radius in the inner regions of the flow \cite{hawley01,hawley02}, rising from $\sim 0.1$ over the bulk of the flow to $\sim 1$ in the plunging region (see, for example, Fig.~4 in \pcite{hawley02} and Fig.~12 in \pcite{mn07}), with the large value of $\alpha$ corresponding to a non-negligible flux of angular momentum being carried out from inside the ISCO as a result of the presence of magnetic fields.   In the present work, we recognize these limitations of the $\alpha$-viscosity formalism and use  $\alpha$ merely as a parameter characterizing the structure of the accretion flow.  We choose its value so that the  spin-up function  that includes the effects of outflows (see \S\ref{sec:spinup})  is comparable to the results seen in MHD simulations.   

Allowing $\alpha$ to vary as a function of black hole spin $j$, we find (see Fig.~\ref{fig:spineq}):
\begin{equationarray}{lcll}
\alpha(j) &=&0.025+0.055 j^2 & \hbox{for } E_{\rm ADAF}=1,\cr
\alpha(j) &=&0.025+0.4 j^4   & \hbox{for } E_{\rm ADAF}=E_{\rm ISCO}.
\label{eq:varalpha}
\end{equationarray}
This dependence of $\alpha$ on $j$ makes qualitative sense if one expects that the angular momentum transport is facilitated by magnetic fields and that the amplitude of these fields ought to grow as the black hole spin is increased.   Such a correlation between magnetic field strength and black hole spin has been noted in numerical simulations \cite{hirose04}. 
Interestingly, as we shall show below, the resulting flow structure, when combined with our jet power model, leads to jet efficiencies comparable to those seen in complex MHD simulations, suggesting that our model may be useful for calculating global properties of AGN systems that are required as input in theories of galaxy, group and cluster formation.  We note that given our usage, our $\alpha$-parameter values cannot be directly compared to physical quantities such as the effective magnetic $\alpha$ of \scite{hawley02} and \scite{mn07}.  In fact, the very concept of $\alpha$-viscosity is not central to our argument.   Any accretion flow with similar structural properties in the inner regions should yield similar results.

The net spin-up function, $s_0^\prime$, for our ADAF model is shown in Figure~\ref{fig:spineq} as cyan curves.  The solid curves show the spin-up function for an ideal ADAF ($E_{\rm ADAF}=1$) while the dashed curves show the spin-up function for the case $E_{\rm ADAF}=E_{\rm ISCO}$, as suggested by numerical simulations. The curves in the left plot are computed using $\alpha(j;E_{\rm ADAF}=1)$  while those in the right plot make use of $\alpha(j;E_{\rm ADAF}=E_{\rm ISCO})$.  We note that in calculating the spin-up rate for an ADAF, we do not include a spin-down term due to radiation swallowing since radiation is not a factor in the case of pure advection flows.    However, we computed the $s_{\rm rad}$ term out of curiosity and found it to be small for all $j$; including it makes no noticeable difference to our results. For the case $E_{\rm ADAF}=E_{\rm ISCO}$ ($E_{\rm ADAF}=1$), the spin-up curve is positive for $j < 0.96$ ($0.97$) and slowly spinning black holes will be spun up as accretion proceeds.   For higher values of $j$, the spin-up curve is negative and the accretion from the ADAF will spin down rapidly whirling black holes. The cross-over point between these two regimes demarcates the equilibrium black hole spin value for accretion via an ADAF.  Specifically, the equilibrium spin is $j=0.96$ for $E_{\rm ADAF}=E_{\rm ISCO}$ and $j=0.97$ for $E_{\rm ADAF}=1$.   The lower equilibrium spin value relative to the thin disk case is the consequence of the accreting fluid's specific angular momentum being sub-Keplerian.

\subsection{Effects of jet production on spin-up}\label{sec:spinup}

We now consider what happens when a jet is launched in conjunction with accretion onto a black hole.  There is a considerable body of work indicating that the launching of jets is most efficient when the accretion flow is advection-dominated (e.g. \pcite{meier01,churazov05}) and that jet production is suppressed in  thin  disks \cite{livio99,meier01,maccarone03}.    We will therefore use the ADAF model described previously as our working platform. Jets are, however, inherently magnetohydrodynamic phenomena \cite{BZ77,BP82,punsly90,meier99,meier01} and our ADAF model does not treat such effects.   As a workaround, we follow the approach outlined by \scite{nemmen07} in developing a model for jet power and  graft it onto our ADAF model.   This approach implicitly assumes the basic structure of the accretion flow will not be significantly altered by the implied presence of strongly ordered magnetic fields in the inner regions.   Recent numerical simulation results of \scite{beckwith08} seem to support this. 

The jet model that we use is described in Appendix~\ref{app:jet}.  The  basic physical idea underlying our model  is that the electromagnetic flux and plasma outflow that comprise the jet are due to a rotating helical tower of magnetic field lines engendered by the  differential frame dragging of a preferentially poloidal field anchored on the event horizon, as well as the combined effect of differential rotation of the plasma in the body of the disk and differential frame-dragging of the plasma inside the black hole's ergosphere on the poloidal field lines anchored in the accretion flow (see \pcite{hirose04,beckwith08}).   In this model,  the jet therefore draws its energy from the energy released by accretion as well as  the rotational energy of the black hole itself, with the contribution from the latter dominating when the black hole is spinning close to its maximal rate.

The fact that jets launched from the black hole-circumnuclear ADAF disk system can derive some or all of their power from the rotational energy of the black hole suggests that they will exert a braking torque on the black hole. Using the irreducible mass of the black hole (\pcite{MTW}; eqn.~33.58):
\begin{equation}
M_{\bullet,\rm irr}={M_\bullet \over 2} \left[ (1+\sqrt{1-j^2})^2+j^2 \right]^{1/2},
\label{eq:Mirr}
\end{equation}
we can derive how the spin $j$ changes due to jet braking. From eqn.~(\ref{eq:Mirr}) we find
\begin{equation}
{\d j \over \d M_\bullet} = {4 \over M_\bullet} \left({M_{\bullet,\rm irr}\over M_\bullet}\right)^2 {\sqrt{1-j^2}\over j}.
\end{equation}
The spin-down rate is then
\begin{equation}
{\d j \over \d t}  = -{\d j \over \d M_\bullet} {\left[f_{\rm BH,\bullet} P_{\rm disk, jet} + P_{\rm BH, jet}\right]\over \c^2},
\end{equation}
and the corresponding jet spin-down parameter is
\begin{eqnarray}
s_{\rm jet} & \equiv & {M_\bullet \over \dot{M}_{\bullet,0} } {\d j \over \d t}\cr
 & =  &- \left[ (1+\sqrt{1-j^2})^2+j^2 \right] {\sqrt{1-j^2}\over j}\cr
 &     &\qquad\qquad\quad\quad\times   \biggl[f_{\rm BH,\bullet} P_{\rm disk, jet} + P_{\rm BH, jet}\biggr].
\end{eqnarray}
In the above equations, $P_{\rm disk, jet}$ and $P_{\rm BH, jet}$ correspond to the jet power associated with the unbounded outflows engendered by magnetic fields anchored in the rotating flow and the black hole event horizon, respectively, and $f_{\rm BH,\bullet}$ is the fraction of the disk jet power that is extracted from the black hole's rotational energy.


\begin{figure}
\psfig{file=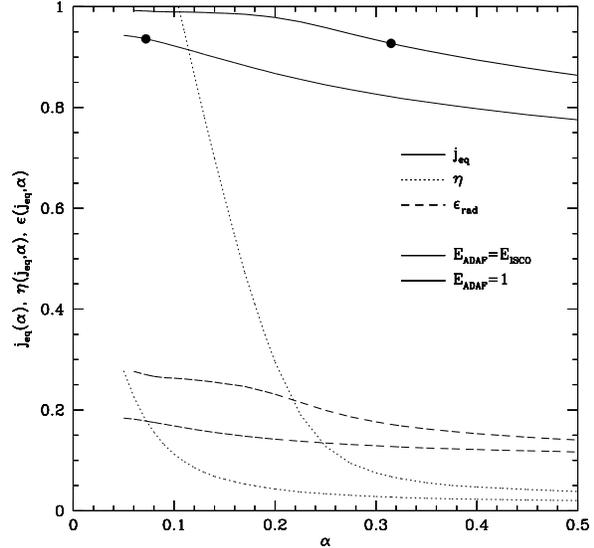,width=80mm}
\caption{The equilibrium spin, $j_{\rm eq}(\alpha)=j(s_{\rm net}=0;\alpha)$, as a function of $\alpha$ for the $E_{\rm ADAF}=1$ case is shown by the thick solid curve while that for $E_{\rm ADAF}=E_{\rm ISCO}$ case is shown as thin solid curve. The dots show the location of the equilibrium spins.  The corresponding maximum radiative efficiency of the black hole, $\epsilon_{\rm rad}[j_{\rm eq}(\alpha)]$, is shown by the dashed lines while the jet efficiency, $\eta=P_{\rm jet}/\dot{M}_{\bullet,0}\c^2$ is shown by the dotted lines.}
\label{fig:jeq}
\end{figure}

The disk and BH jet power are given by (see Appendix~\ref{app:jet} for details) 
\begin{eqnarray}
 P_{\rm disk,jet}&=&{3 \over 80} (1-\beta) g^2 r^2 \dot{M}_{\bullet,0}{\gamma_r^2\gamma_\phi^2\over {\mathcal A} V {\mathcal
 H}} \sqrt{{1-V^2\over {\mathcal D}}}\widetilde{T} \cr
 & & \times\left( {{\mathcal L}_{\rm ADAF}\over r^2\gamma_r\gamma_\phi}\sqrt{{{\mathcal D}\over{\mathcal A}^3}}+{2j\over {\mathcal A}r^3} \right)^2 \cr
 & & \times \left( {2 j \beta_\phi \over r^2} + \sqrt{{\mathcal D}} \right)^2  ,
\end{eqnarray}
and
\begin{eqnarray}
 P_{\rm BH,jet} &=& {3 \over 80} (1-\beta) g^2 r^2 \dot{M}_{\bullet,0} {\gamma_r^2\gamma_\phi^2\over {\mathcal A} V {\mathcal H}} \sqrt{{1-V^2\over {\mathcal D}}}\widetilde{T} \left( {2j\over {\mathcal A}r^3} \right)^2 \cr
 & & \times \left( {2 j \beta_\phi \over r^2} + \sqrt{{\mathcal D}} \right)^2,
\end{eqnarray}
where all quantities are evaluated at the ISCO for the disk jet and at the static limit for the BH jet, the relativistic boost factors $\gamma_r$ and $\gamma_\phi$ are defined by eqns.~(\ref{eq:gammar}) and (\ref{eq:gammaphi}) respectively, $\beta_\phi = \sqrt{1-1/\gamma_\phi^2}$, the metric factors ${\mathcal A}$ and ${\mathcal D}$ are defined by eqns.~(\ref{eq:metA}) and (\ref{eq:metD}) respectively, $V$ is the radial velocity of the flow given by eqn.~(\ref{eq:radvel}), ${\mathcal L}_{\rm ADAF}$ is given by eqn.~(\ref{eq:angL}), $\widetilde{T}$ is a dimensionless temperature given by eqn.(\ref{eq:temp}), and $g$ is a magnetic field enhancement factor due to shearing (see Appendix~\ref{app:jet}) and is given by $g=\exp(\omega \tau)$, where $\omega$ is the angular velocity of spacetime rotation given by eqn.~(\ref{eq:framedrag}) and $\tau$ is a characteristic timescale available for field enhancement given by eqn.~(\ref{eq:timescale}).

The BH jet power vanishes as the black hole spin approaches zero.   In other words, $P_{\rm BH,jet}=0$ for the Schwarzchild black hole.   However, the unbounded flows in the numerical simulations \cite{deV03,vhkh05,hawley06} do not vanish for this case, indicating that the accretion flow itself also makes a contribution. The disk jet power does not go to zero as the black hole spin goes to zero, but it is not independent of the black hole spin either.  It grows with increasing spin.   The effects of frame-dragging of the accretion flow within the ergosphere of a spinning black hole enhances the power of the disk jet by a factor of $g^2$ (see Appendix~\ref{app:jet}). We use this to estimate the fraction of the power that must come from the accretion power.  Since $g=1$ in absence of frame-dragging, we adopt $f_{\rm BH,\bullet}=1-g^{-2}$ as a reasonable approximation for the fraction of the disk jet power that is extracted from the black hole's rotational energy.  This approximation states that regardless of the value of the black hole spin, accretion will always provide some amount of power to the outflows.   For a Schwarzchild black hole, the disk (and total) jet power is then driven entirely by accretion power ($\dot{M}\c^2$) while for a maximally spinning Kerr black hole only a fraction of a percent of the disk and total jet power is extracted from accretion power, the rest being extracted from the black hole spin. Specifically,  72.8\% of the disk jet power (about 93\% of the total jet power) is drawn from the black hole's rotational energy for $j=0.8$, rising to 92.9\%  (96.6\%) at $j=0.9$.  We note that the results that we present in this paper are not sensitive to the details of this approximation providing that, for $j$ close to unity, the majority of the jet power is drawn from the black hole's rotational energy. As we will show below, this is indeed the case.
 
\begin{figure*}
\centering 
\includegraphics[angle=270,width=180mm]{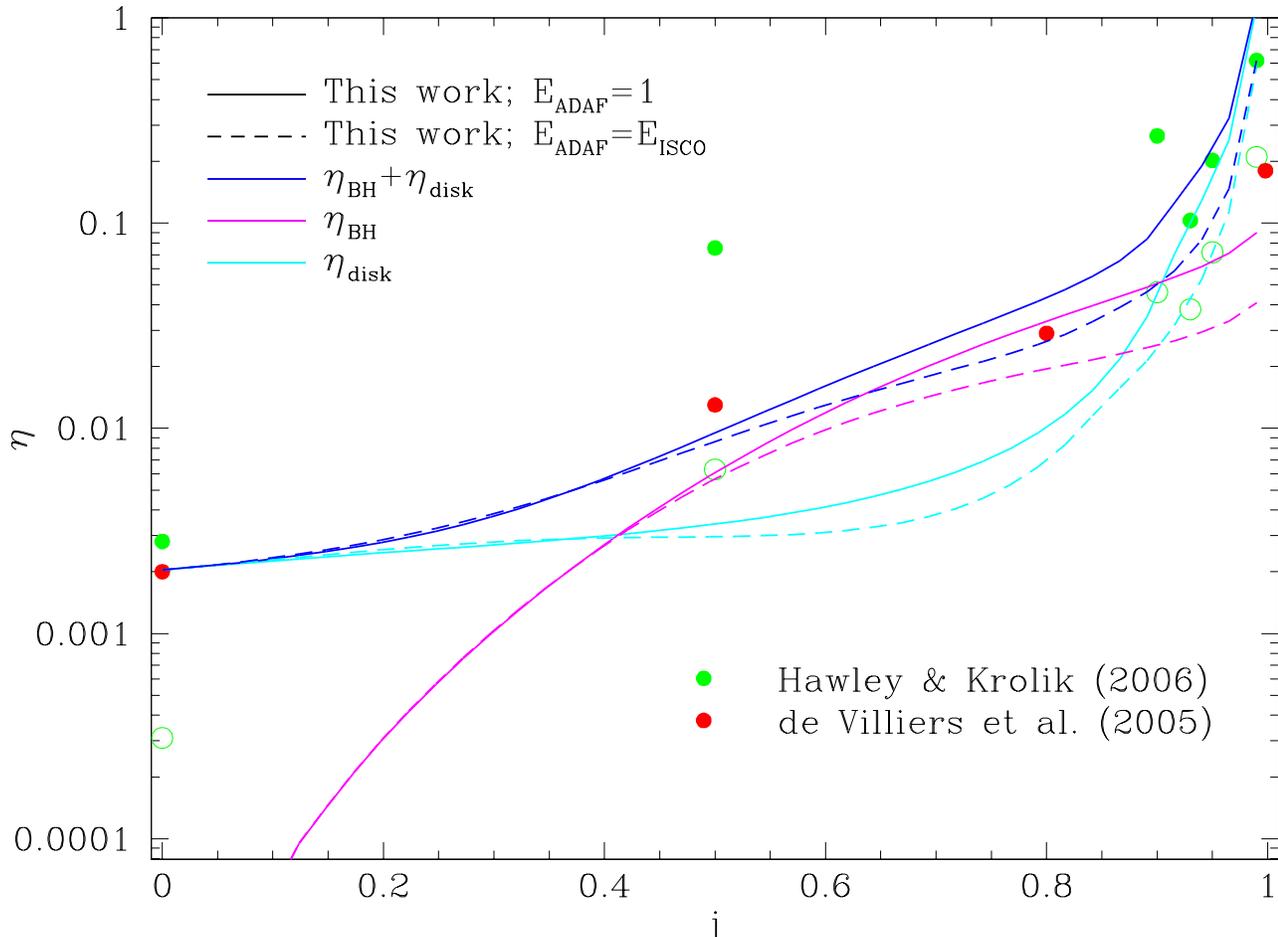} 
\caption{The jet efficiency, $\eta=P_{\rm jet}/\dot{M}_{\bullet,0}\c^2$, predicted by our ADAF model with  $\gamma=1.444$ is shown as a function of $j$ (lines). The solid line shows the results for our $E_{\rm ADAF}=1$ model while the dashed line shows the results for our $E_{\rm ADAF}=E_{\rm ISCO}$. Magenta lines show the black hole (electromagnetic) jet efficiency, cyan lines show the disk  (electromagnetic and matter) jet efficiency, and blue lines show the sum of the two. Estimates of $\eta_{\rm tot}$ from the numerical simulations of \protect\scite{hawley06} and \protect\scite{vhkh05} are shown by green and red circles respectively, with the filled points corresponding to the efficiency factor of the  total jet power and the open circles corresponding to the efficiency factor of the Poynting-flux component. }
\label{fig:eta_jet}
\end{figure*}

The net spin-up function for $E_{\rm ADAF}=1$ and $E_{\rm ADAF}=E_{\rm ISCO}$ ADAFs with jet outflows are plotted in Fig.~\ref{fig:spineq} as red curves.  Our results display the same qualitative behaviour as the \scite{krolik05} and \scite{gsm04} simulation results, which we also show.   This correspondence with the simulation results holds even if the $\alpha$ parameter is treated as a constant.  Our results, therefore, make a strong prediction that there should be an equilibrium value of black hole spin where we may expect to see the spin distribution be truncated (if black holes spin up from $j=0$ via an ADAF).

For $\alpha(j)$ as described in eqn.~(\ref{eq:varalpha}), the spin-up functions for ADAFs with and without jets are nearly identical at low spin values ($j\lsim 0.8$).  As noted previously, the jet power at low spin values is largely drawn from the accretion flow and is relatively low.  At high spin values, however, the spin-up rate is noticeably lower once the jet power is taken into account, with the effect being more pronounced for the ideal ADAF case.    Note that for $j < 0.92$, $s_{\rm net}$ is positive and low spinning black holes will be spun up as accretion proceeds, even if the accretion is accompanied by outflows.   For $j>0.94$, however, $s_{\rm net}$ is negative;  the combination of jets and angular momentum flux from the ADAF will spin down rapidly whirling black holes. The cross-over point between these two regimes demarcates the equilibrium black hole spin value for accretion via an ADAF.  Specifically, the equilibrium spin is  $j=0.93$ for $E_{\rm ADAF}=1$ and $j=0.92$ for $E_{\rm ADAF}=E_{\rm ISCO}$   The $\alpha$ value corresponding to the equilibrium spin value is 0.073 if $E_{\rm ADAF}=1$ and 0.317 if $E_{\rm ADAF}=E_{\rm ISCO}$.  

The equilibrium spin depends on the dimensionless parameters $\alpha$, $\gamma$ and $f$ that describe the structure of the ADAF.  Setting aside the fact that we have chosen our $\alpha(j)$ so that the resulting spin-up function agrees with simulation results, one can ask how sensitive the equilibrium value is to the value of $\alpha$ at the point where $s_{\rm net}=0$.    In Fig.~\ref{fig:jeq} we plot the equilibrium spin $j(s_{\rm net}=0)$ as a function of $\alpha$ (with fixed $f=1$ and $\gamma=1.444$).   As the plot illustrates, a decrease in the value of $\alpha$ results in a higher value of $j_{\rm eq}$, and vice versa; the dependence, however, is relatively weak.    Over the range, $\alpha=0.05$--$0.5$, the change in the equilibrium spin value is of order $10\%$.      

Fig.~\ref{fig:jeq} also shows the run of  jet efficiency, defined as $\eta=P_{\rm jet}/\dot{M}_{\bullet,0}\c^2$ corresponding to the equilibrium spin value, where, as noted previously, $\dot{M}_{\bullet,0}$ is the rate of rest mass accreting onto the black hole\footnote{The jet efficiency, $\eta$, defined in this way can exceed unity since the jet can draw power not only from the accretion flow, but also from the black hole spin.}.  For equilibrium spin value  $j\approx 0.92$, the jet efficiency factor is $\eta=0.06$ and $0.16$ for $E_{\rm ADAF}=E_{\rm ISCO}$ and $E_{\rm ADAF}=1$, respectively.  For values of $\alpha\gsim 0.3$, the jet efficiency parameter is nearly constant.   This is because the corresponding equilibrium spin values are not so high and the jet power comes largely from accretion.  The equilibrium spin value simply represents the maximum spin state that accretion can drive the system into.  The decline in equilibrium spin (and consequently jet efficiencies) with increasing $\alpha$ is due to the reduction in the angular momentum of the ADAF material.   For $\alpha\lsim 0.2$, jet efficiency rises steeply with decreasing $\alpha$.   Even though the magnitude of the equilibrium spin value has not increased by much, the system is now moving to a state where the jet power is coming to be dominated by the rotational energy of  the black hole.   The equilibrium spin values seen in numerical simulations suggest that those systems are close to this latter state.   In this state,  the angular momentum being carried by the jet is not negligible and the equilibrium spin value is truly a result of a competition between spin-up due to accretion and spin-down due to the jet.   

For completeness, we also show in Fig.~\ref{fig:jeq}, the maximum radiative efficiency of the black hole, $\epsilon_{\rm rad}(s_{\rm net}=0)=1-E_{\rm ISCO}[j(s_{\rm net}=0)]$ as a function of $\alpha$.  For both of our fiducial models, we find $\epsilon_{\rm rad}(s_{\rm net}=0)\approx 0.16$ and like the equilibrium spin, the magnitude of $\epsilon_{\rm rad}$ is only weakly dependent on $\alpha$.   We would be remiss if we did not mention that this radiative efficiency does not, of course, apply to an optically thin ADAF.  However, we would expect this efficiency to be relevant to an accretion system that has been spun up by an ADAF and then transitions to a radiatively efficient thin-disk mode but has not yet been in this new state long enough to alter the black hole spin.  For completeness, we note that in general, the radiative efficiency of an AGN can be modeled approximately as 
\begin{equation}
\epsilon_{\rm ADAF}\approx\epsilon_{\rm rad}(j)\times \left\{ 
\begin{array}{l l} \left(\dot{m}/\dot{m}_{\rm cr}\right), & \quad \mbox{if  $\dot{m} < \dot{m}_{\rm cr}$}\\
                               1,  & \quad \mbox{if  $\dot{m} > \dot{m}_{\rm cr}$}\\ \end{array} \right.
\end{equation}
where $\dot{m}$ is the Eddington-scaled mass accretion rate and $\dot{m}_{\rm cr}=0.03$ is the rate above which the accretion flow becomes radiatively efficient \cite{merloni08}.

In Fig.~\ref{fig:eta_jet}, we show the jet efficiency as a function of $j$ for our fiducial $E_{\rm ADAF}=1$ and $E_{\rm ADAF}=E_{\rm ISCO}$ (solid and dashed blue lines), and compare these to the available results from numerical simulations.   The magenta curves show the black hole jet efficiency,  which vanishes at zero black hole spin; the cyan curves show the disk jet efficiency, which asymptotes to $\eta_{\rm disk}\approx 0.002$ as $j$ vanishes; and the blue curves show the efficiency factor for the total jet power.   The simulation results are shown as green and red circles.  The filled circles show the efficiency of the combined power in electromagnetic and matter\footnote{Like both \protect\scite{hawley06} and \protect\scite{vhkh05}, we remove the rest mass contribution to the matter outflow.} outflows, and the open circles show the efficiency associated with the electromagnetic jet.    While there remain significant differences between the various numerical calculations of $\eta$ (e.g. there is an order of magnitude difference in the jet power at $j=0.5$ as determined by the two different calculations),  there is a clear and rapid increase in jet efficiency with rising $j$.  This is reproduced by our model, which is in reasonably good agreement with the simulation results.  Specifically, the simulation results indicate that even at $j=0$, there are outflows from accretion flow-black hole system.    Our model reproduces this non-zero jet efficiency at zero spin and indicates that it is powered entirely by the rotation of the accreting plasma.    At intermediate spins ($j\approx0.5$---0.8), the black hole jet dominates.   At high spins ($j>0.9$), the simulation results show a rapid rise in power.   Our black hole jet model, which is based on the Blandford-Znajek model \cite{BZ77}, does not rise as rapidly and this has been one of the many criticisms of this particular model.   Our total jet power, however, does show a steep rise and the results are consistent with the simulation results.   The steep rise is driven by a sharp increase in the disk jet power.    In our model, we cannot distinguish whether this steep rise in the disk jet power is due to increase in the kinetic flux of the matter outflow, the electromagnetic flux, or both.  However, an analysis of the \scite{hawley06} numerical simulations has led \scite{punsly07} to conclude that for high spins, the effects of frame-dragging can not only result in disk jet that is Poynting-flux dominated but 
also enhance its power so that it dwarfs the black hole jet.

\section{Discussion}\label{sec:discuss}

Jets appear to be a ubiquitous phenomenon in accreting black hole systems, with both the accretion flow as well as the black hole's rotational energy contributing to the jet power.  The question of how much energy can be extracted from an accretion flow-rotating black hole system has been addressed previously by a number of groups, the most recent being \scite{nemmen07}.  The jets, however, also carry away angular momentum from the system and most of these studies have not considered the implications of this angular momentum extraction.  Since jets emitted by systems with rotating black holes derive a fraction of their power from the spin of the accreting black hole \cite{meier01,nemmen07}, they must exert a braking torque on the black hole.   Jets, therefore, must play a central role in limiting the spin-up of black holes.    

In a nutshell, the spin equilibrium is the result of the coupling between the black hole, the accretion disk and the jet outflows mediated by magnetic fields.   The detailed mechanism through which the magnetic fields couple the rotational energy of the black hole to the outflows is a topic of a number of on-going analytic and numerical investigations, and we presented a summary of some of the recent insights and thinking about the origin of the jet phenomenon in \S1.
In this paper, we have described a simple calculation that attempts to capture some of the aspects of the complex process summarized above in the form of a highly simplified model. The resulting  calculation, based on the ADAF model summarized in Appendix~\ref{app:adaf} and jet model summarized in Appendix~\ref{app:jet}, indicates that any black hole which has undergone significant accretion (i.e. more than doubling its mass) from an ADAF-type flow will have a spin of approximately $j\approx 0.92$ providing it remains undisturbed afterwards (i.e. no mergers or subsequent accretion by other means).  Moreover, the predicted equilibrium spin should be independent of the initial spin state. For example, a black hole that begins accreting from an ADAF with an initial $j>j(s_{\rm net}=0)$ (where the jet power is very large) will be rapidly spun-down to the equilibrium value.  Essentially, any black hole driving a sufficiently powerful jet will reach an equilibrium spin $j<1$.   This fundamental result --- that black holes driving powerful jets must experience a braking torque which plays a central role in limiting their spin --- is independent of the details of the model that we employ.

The  calculation we present  is in excellent agreement with results from MHD simulations of accreting black hole systems.   Specifically, we show that by using a model that matches the spin-up function measured in the numerical MHD simulations, we are able to self-consistently explain the jet efficiencies measured in the same simulations.   This dual agreement is a pleasant surprise, given the simplifications and approximations inherent in our quasi-analytic approach and suggests that our model represents a convenient useful description of ADAF inflow -- jet outflow that can be easily incorporated in semi-analytic modeling as well as numerical simulations focusing on the formation and evolution of galaxies, groups and clusters of galaxies.  Numerous studies have shown that AGN feedback, both kinetic and radiative, is critical for preventing the formation of overluminous, much too blue, massive galaxies \cite{hopkins05,croton06,bower06,somerville08} and for tempering massive cooling flows that ought to have been present in at least 30\% of galaxy clusters \cite{mccarthy08,bildfell08}, but which the observations suggest are much weaker \cite{Ka01,P01}.  The nominal explanation is that radiative energy loss that underlies the cooling flow phenomena is being partially offset by heating by outflows from supermassive AGNs in cluster galaxies.

Recently, \scite{allen06} analyzed X-ray and optical data from set of nine nearby, X-ray luminous giant elliptical galaxies that show evidence of jet outflow-related activity and found a remarkably tight correlation between the estimated Bondi accretion rate\footnote{Given a distribution of gas about a central black hole,  the most simple configuration describing the accretion of the gas onto the the black hole is the Bondi flow model \cite{bondi52}, which assumes a non-luminous central source and a spherically symmetric flow with negligible angular momentum.    The resulting Bondi accretion rate can be written as $\dot{M}_{\rm Bondi}=\pi \lambda c_s \rho r^2_A$, where $r_A=2GM_{\bullet,0}/c_s^2$ is the accretion radius, $G$ is the gravitational constant, $M_{\bullet,0}$ is the black hole mass, $c_s$ is the sound speed of the gas at $r_A$, $\rho$ is the density of gas at $r_A$ and $\lambda$ is a numerical coefficient that depends on the adiabatic index of the gas.} and the estimated jet powers in these systems of the form:   $P_{\rm jet}=\eta_{\rm Bondi} \dot{M}_{\rm Bondi} c^2$, where $\eta_{\rm Bondi}\approx 0.02$.  The results are consistent with those expected from the ADAF-jet model of the kind presented in the paper especially if one allows for the fact that not all of the mass flow at the Bondi radius will actually accrete onto the black hole.   For example, if $\dot{M}_{\bullet,0} = f \dot{M}_{\rm Bondi}$ where $f<1$, then the jet efficiency factor, as defined and used in this paper, is $\eta = \eta_{\rm Bondi}/f$.   A careful analysis of the constraints on $\eta$  imposed by the trend found by  \scite{allen06} led \scite{nemmen07} to conclude that the resulting black holes must be spinning rapidly (i.e. $j \gsim 0.8$).   An examination of Fig.~\ref{fig:eta_jet} shows that the use of our model leads to very similar conclusions.   More importantly, though, 
Fig.~\ref{fig:eta_jet} shows that jet efficiency is a strong function of the black hole spin.   Over the range, $0.8 < j < 1.0$, $\eta$ varies by approximately two orders of magnitude.  This suggests that if the spin of the black holes in massive elliptical galaxies is unconstrained and any value between 0.8 and 1 is equally likely,   then one would expect to find $P_{\rm jet}$ varying by as much as two orders of magnitude for a given value of $\dot{M}_{\rm Bondi}$ and therefore, the \scite{allen06}  plot should have resembled a scatterplot.  The fact that it does not indicates that the black hole spins are not unconstrained but rather, span a tight distribution.    We argue, based on the findings presented in this paper, that the most natural value for the black hole spins to cluster about is the equilibrium spin value of $j\approx 0.92$.   

Constraining the black hole spins to $j\approx 0.92$, with the corresponding jet efficiency of $\eta\approx 0.06$, further implies that $\sim 30$\% of the material Bondi flow accretes on the black hole.   This surprisingly large fraction indicates that the black hole is very efficient at capturing material within its Bondi radius, which for the systems in question is approximately $10$-$20$ parsecs.

Finally, we emphasize that the calculation arguing for the existence of a spin equilibrium at $j\sim 0.92$ is predicated on the accretion disk-black hole system driving powerful outflows from an ADAF.  Black holes accreting via thin accretion disks will behave very differently.  The jet power produced by a thin disk is around three to four orders of magnitude smaller than that from an ADAF for $j \approx 1$ \cite{meier01}.  The maximum $|s_{\rm jet}|=4.1\times 10^{-5}$ for a standard thin disk  occurs at $j=0.94$.  This is much less than $|s_{\rm rad}|$ which is approximately $0.016$ at the same $j$.  The jet will have little impact on the black hole spin and our model predicts that  a black hole spun up by accretion from a thin disk should reach the limiting spin of $j=0.998$ found by \scite{thorne74}.

\section*{Acknowledgments}

We are grateful to Charles Gammie, Julian Krolik, David Meier and the anonymous referee for insightful discussions and clarifications and to Rodrigo Nemmen for making available to us his model for the jet power in advance of publication. AJB acknowledges support from the Gordon and Betty Moore Foundation and would like to acknowledge the hospitality of the KITP at the University of California, Santa Barbara where part of this work was completed. AB acknowledges support from the Leverhulme Trust (UK) and NSERC (Canada), and is deeply appreciative of the hospitality shown to him by Richard Bower, Carlos Frenk, Joe Silk, and more generally by the Institute of Computational Cosmology (University of Durham) and the Department of Astrophysics at the University of Oxford during his tenure there as the Leverhulme Visiting Professor. This research was supported in part by the National Science 
Foundation under Grant No. NSF PHY05-51164 .We thank Alejo Martinez-Sansigre for bringing errors in the original version of this work to our attention.

\onecolumn
\appendix

\section{ADAF Flow onto a Kerr Black Hole}\label{app:adaf}

In this Appendix, we describe our model for an advection-dominated accretion flow onto a Kerr black hole.    Our model is based on the results of \scite{popgam98}, who numerically solved the equations describing the radial structure of steady state, optically thin, advection-dominated accretion flow in a Kerr metric (see also, \pcite{gampop98,manmoto00}).  The accreting fluid is assumed to be a mixture of a frozen-in,  isotropically tangled, turbulent magnetic field and ionized plasma, and the key parameters that determine the structure of the ADAF are (i) the dimensionless black hole spin parameter $j$,  (ii) the adiabatic index $\gamma$ quantifying the strength of the magnetic fields in the flow, (iii) the viscosity parameter $\alpha$ governing the transport of angular momentum in the flow, and (iv) the advected fraction of the dissipated energy $f$.  

We make a number of simplifying approximations and assumptions.  The first of these is that we will assume, like \scite{gampop98}, that the angular momentum of the accreting fluid is aligned with the angular momentum of the black hole.   Additionally, we will also focus our attention specifically on the structure of the flow close to the equatorial plane.   Also, as noted in the text, we shall set $f=1$; that is, we shall only consider pure advection flows.  We have discussed our treatment of the viscosity parameter $\alpha$ in the text.   The adiabatic index of the fluid will depend on the relative contributions of the thermal and magnetic energy densities to the total internal energy density, $U_{\rm tot} = U_{\rm mag}+ U_{\rm gas}$, of the fluid.   If the magnetic field contributes energy density $U_{\rm mag} =B^2/8\pi$ to the admixture, then the corresponding isotropic magnetic pressure is $P_{\rm mag} = U_{\rm mag}/3$ \cite{narayan98}.   Assuming that the magnetic pressure contributes a constant fraction of the total pressure, $P_{\rm mag}=(1-\beta)P_{\rm tot}$, it is straightforward to show that (c.f., \pcite{esin97}):
\begin{equation}
\gamma = {8-3\beta \over 6-3\beta }.
\label{eq:gamma}
\end{equation}  
Since $0 \leq \beta \leq 1$, $\gamma$ is constrained to range within $[4/3,\,5/3]$.  For a mixture in which the magnetic field is in equipartition (i.e.~$U_{\rm gas}=U_{\rm mag}$), $\beta=2/3$ and $\gamma = 1.5$.   (As a point of clarification, we note that all thermodynamic quantities, including magnetic field pressure and energy density, are measured in the fluid's local rest frame.)

In eqn.~(\ref{eq:gamma}), we have assumed that the  ionized plasma can be described adequately as an ideal, classical gas.  Specifically, we ignore any relativistic corrections that may arise at high temperatures.   Finally, in the discussion below,  we shall, for convenience, work in geometric units ($\G=\c=M_\bullet=1$); the units of mass, length and time are $M_\bullet$, $\G M_\bullet/\c^2$, and $\G M_\bullet/\c^3$, respectively.   We also emphasize that \scite{popgam98} (see \pcite{gampop98} for details) do not  take into account the effects of any large-scale magnetic fields that may be present, nor do they treat MHD effects, like the amplification of magnetic fields by dynamo-like processes.  Consequently, neither does the model presented below.      

In the reference frame of an observer who is corotating with the fluid about the black hole at the Boyer-Lindquist radial coordinate $r$, the radial velocity of the accreting material (c.f.~\pcite{popgam98}; Figs.~1--4) is well described by fitting function $V(r)$ where 
\begin{equation}
V(r) = -\sqrt{1-(1-2/r_{\rm eff}+(j/r_{\rm eff})^2)},
\label{eq:radvel}
\end{equation}
where
\begin{eqnarray*}
r_{\rm eff} &=& r_{\rm h}+ \Psi(r) (r-r_{\rm h}),\\
\Psi(r) & = & v_1 v_2 v_3 v_4 v_5, \\
v_1 &=& 9\log(9z), \\
v_2 &=& \exp(-0.66[1-2 \alpha_{\rm eff}]\log[\alpha_{\rm eff}/0.1]\log[z/z_{\rm h}]), \\
v_3 &=& 1-\exp[-z\left\{0.16(j-1)+0.76\right\}], \\
v_4 &=& 1.4+0.29065(j-0.5)^4-0.8756(j-0.5)^2)+(-0.33j+0.45035)[1-\exp(-(z-z_{\rm h}))], \\
v_5 &=& 2.3\exp[40(j-1)]\exp[-15r_{\rm ISCO}(z-z_{\rm h})]+1. \\
\alpha_{\rm eff}& = & \alpha\left[ 1+6.450(\gamma-1.444)+1.355(\gamma-1.444)^2\right].
\end{eqnarray*}
In this series of equations, $z=r/r_{\rm ISCO}$ and $z_{\rm h} =r_{\rm h}/r_{\rm ISCO}$, where $r_{\rm h}(j) = 1+\sqrt{1-j^2}$ is the radius of the event horizon and  $r_{\rm ISCO}(j)$ is the radius of the ISCO (see eqn.~\ref{eq:isco}).

Popham \& Gammie's steady state ADAF is characterized by a constant rest-mass accretion rate, $\dot{M}$.  The combination of this accretion rate and the above velocity can be used, via the continuity equation,  to compute the radial distribution of the rest-mass density, $\rho(r)$:
\begin{equation}
\rho(r)={-\dot{M}\over 4\pi r^2 {\mathcal{H}}V}\sqrt{{1-V^2}\over {\mathcal{D}}},
\end{equation}
where ${\mathcal{H}}(r)$ is the characteristic angular scale of the flow about the equator, which according to \scite{popgam98} is given by
\begin{equation}
{\mathcal{H}}^2 = {\widetilde{T}\over \eta r^2 \nu_{\rm z}^2},
\end{equation}
where $\widetilde{T}$ is the dimensionless temperature defined in eqn.~(\ref{eq:temp}). In these equations, $\eta\equiv (P_{\rm tot} + U_{\rm tot} + \rho)/\rho $ is the relativistic enthalpy of the accreting fluid, 
\begin{equation}
\nu_{\rm z}^2 = {1\over r^4} \left\{j^2 + [1-(j\omega)^2] {\mathcal L}_{\rm ADAF}^2 - [(j\gamma_\phi)^2/{\mathcal A}] (\gamma_r \sqrt{{\mathcal D}})^2 - \gamma_r\sqrt{{\mathcal D}} [2 {\mathcal L}_{\rm ADAF} \omega \gamma_\phi j^2]/\sqrt{{\mathcal A}}\right\},
\end{equation}
is the effective vertical frequency,
\begin{equation}
{\mathcal{D}}(r) = 1-2/r+(j/r)^2
\label{eq:metD} 
\end{equation}
and
\begin{equation}
{\mathcal A} = 1 + {j/ r}^2 + 2 {j^2/ r^3}
\label{eq:metA}
\end{equation}
are relativistic (metric-related) factors, 
\begin{equation}
\gamma_r =   \sqrt{1\over 1-V^2}
\label{eq:gammar}
\end{equation}
and
\begin{equation}
\gamma_\phi  =  \sqrt{1+{{\mathcal L}_{\rm ADAF}^2 \over r^2 {\mathcal A} \gamma_r^2}}
\label{eq:gammaphi}
\end{equation}
are relativistic boost factors associated with the radial and tangential motions of the ADAF fluid with respect to the Boyer-Lindquist coordinates \cite{popgam98},
\begin{equation}
\omega  \equiv  - {g_{\phi t} \over g_{\phi \phi}} = {2 j \over {\mathcal A} r^3 },
\label{eq:framedrag}
\end{equation}
is the angular velocity, in the same coordinate system, corresponding to the  local spacetime rotation (frame-dragging) enforced by the spinning black hole,   ${\mathcal L}_{\rm ADAF}$ is the  angular momentum of the accreting fluid, and $P_{\rm tot}$ is the fluid's total (thermal plus magnetic) pressure.

\scite{popgam98} do not give the total pressure explicitly.   Rather, they plot the generalized dimensionless temperature of the accreting fluid, $\widetilde{T}(r,j,\alpha,\gamma)\equiv P_{\rm tot}/\rho$, which can then be used to compute the total pressure.   This dimensionless temperature is well described by the fit:
\begin{equation}
\widetilde{T}(r,j,\alpha,\gamma) = 0.31 {(1+[t_4(j)/r]^{0.9})^{t_2(\gamma)+t_3(\alpha)} \over [r-t_5(j)]^{t_1(\gamma)}},
\label{eq:temp}
\end{equation}
where
\begin{eqnarray*}
t_1(\gamma) &=& -0.270278\gamma+1.36027, \\
t_2(\gamma) &=& -0.94+4.4744(\gamma-1.444)-5.1402(\gamma-1.444)^2, \\
t_3(\alpha) &=& 0.84\log_{10}\alpha+0.919-0.643\exp(-0.209/\alpha), \\
t_4(j) &=& [0.6365r_{\rm ISCO}(j)-0.4828][1.0+11.9\exp\left\{-0.838r_{\rm ISCO}^4(j)\right\}], \\
t_5(j) &=& 1.444\exp[-1.01r_{\rm ISCO}^{0.86}(j)]+0.1.
\end{eqnarray*}
The  relativistic enthalpy can also be expressed in terms of the dimensionless temperature: 
\begin{equation}
\eta\approx 1+ {\gamma\over \gamma -1}\widetilde{T}(r,j,\alpha,\gamma).
\end{equation}
Here $\gamma$ is the adiabatic index of the accreting fluid.   

The one remaining quantity that appears in the above equations but has yet to be defined ${\mathcal L}_{\rm ADAF}$, the specific angular momentum of the accreting fluid.   We determine this from the product $\eta {\mathcal L}_{\rm ADAF}$ that \scite{popgam98} plot (c.f.~ their Figs.~1--4).   Our fit for this quantity is:
\begin{equation}
\eta {\mathcal L}_{\rm ADAF} =  (\eta {\mathcal L}_{\rm ADAF})_2+[(\eta {\mathcal L}_{\rm ADAF})_1+10^{(\eta {\mathcal L}_{\rm ADAF})_6}] [1.15-0.03(\log_{10}\alpha+3)^{2.37}],
\label{eq:angL}
\end{equation}
where
\begin{eqnarray}
(\eta {\mathcal L}_{\rm ADAF})_1 &=& 0.0871 r_{\rm ISCO}-0.10282, \\
(\eta {\mathcal L}_{\rm ADAF})_2 &=& 0.5-7.7983(\gamma-1.333)^{1.26},  \\
(\eta {\mathcal L}_{\rm ADAF})_3 &=& 0.153 (r_{\rm ISCO}-0.6)^{0.30}+0.105, \\
(\eta {\mathcal L}_{\rm ADAF})_4 &=& (\eta {\mathcal L}_{\rm ADAF})_{3} (0.9\gamma-0.2996)(1.202-0.08[\log_{10}\alpha+2.5]^{2.6}), \\
(\eta {\mathcal L}_{\rm ADAF})_5 &=& -1.8\gamma+4.299-0.018+0.018(\log_{10}\alpha+2)^{3.571}, \\
(\eta {\mathcal L}_{\rm ADAF})_6 &=& (\eta {\mathcal L}_{\rm ADAF})_4 \left\{[(0.14 (\log_{10}r)^{(\eta {\mathcal L}_{\rm ADAF})_5}+0.23)/(\eta {\mathcal L}_{\rm ADAF})_4]^{10}+1\right\}^{0.1}.
\end{eqnarray}

These relations cumulatively completely specify the basic structure of a ``simple'' ADAF.

\section{Jet Power from an ADAF}\label{app:jet}

In this Appendix, we describe a model for computing the jet power arising from our ADAF, including the jet launched from the disk and the jet launched from the black hole. We will assume that these two jets arise from the same physical mechanism, differing primarily in the radii from which they originate. There is a considerable body of work indicating that the launching of jets is most efficient when the accretion flow is geometrically thick, advection-dominated flow (e.g., \pcite{meier01,churazov05}) and least efficient when it proceeds via a geometrically thin accretion disk normally associated with radiatively efficient AGNs  \cite{livio99,meier01,maccarone03}.  While a detailed understanding of the extragalactic AGN jet phenomena remains elusive, the combination of physically insightful analytic studies by a number of authors over the past three decades and recent sophisticated general relativistic, magnetohydrodynamic (MHD) numerical simulations are beginning to yield important insights. There is now a general consensus that jets are fundamentally MHD events.  Since our ADAF model described in Appendix~\ref{app:adaf} does not treat MHD effects, we will develop a  separate model for jet power that can then be coupled to our ADAF model.

The jet model that we  adopt follows the construction outlined by \scite{nemmen07}, which itself is an improved version of the scheme first proposed by \scite{meier01}.  We recognize that this disjointed approach is not fully self-consistent, but we accept this and other associated limitations in favour of a simple, easy-to-use, model.  That our model jet results agree reasonably well with those found in MHD simulations gives us a measure of confidence.   As in Appendix~\ref{app:adaf}, the description below will make use of  geometric units where $\G=\c=M_\bullet=1$, and we will, as before, seek to keep the model simple by adopting as input ADAF properties computed in the equatorial plane of the flow.

As our starting point, we assume that both the disk and the black hole jet powers are given by \cite{meier01}:
\begin{equation}
P_{\rm jet} = {1\over 32} \left[ r^2 B^{\rm LS}_{\rm pol, \infty}(r) \Omega_\infty(r) \right]^2,
\label{eq:Pjet}
\end{equation}
where all quantities are evaluated at a characteristic radius associated with the region within which the jet forms. For the jet launched from the black hole, we will assume a characteristic radius equal to the static limit (i.e. the edge of the ergosphere) in the disk plane, $\theta=\pi/2$, i.e.~$R_{\rm static}=1+\sqrt{1-j^2\cos^2\theta}=2$ in gravitational units.   For the disk jet, we identify the characteristic radius with $R_{\rm ISCO}$ under the assumption that the jet originates from the region interior to the inner edge of the ADAF and that this inner edge coincides with $R_{\rm ISCO}$. We acknowledge that there is an ambiguity associated with this identification.  As discussed by \scite{kh02} and \scite{wm03}, there is no reason for the inner edge to occur precisely at the ISCO and, furthermore, the location of the ``inner edge'' will depend on the physical property that defines that edge.  The various simulation studies suggest that the launch region could be a factor of 2--3 inside the ISCO.  Under our scheme, a smaller characteristic radius would lead to slightly higher jet power.   For example, at $j=0.9$, the jet power is enhanced by a factor of two if the innermost edge occurs half-way between the ISCO and the event horizon.   We consider such order unity uncertainties as an acceptable compromise. 

In eqn.~(\ref{eq:Pjet}), $B^{\rm LS}_{\rm pol, \infty}$ is the large-scale poloidal magnetic field, as seen by an observer at infinity in the Boyer-Lindquist coordinate system and $\Omega_\infty$ is the angular velocity of the rotating magnetic fields (and the plasma), as seen by the same observer.    For the disk jet, this angular velocity takes into account the fact that the fluid itself is rotating as well as the effects of frame-dragging if the black hole itself is spinning; the black hole jet depends only on the angular velocity associated with the frame-dragging enforced by the spinning black hole:
\begin{equation}
\Omega_\infty(r) = \left\{ \begin{array}{ll}
\widetilde{\Omega}(r) + \omega(r) & \hbox{for disk jet,} \\
\omega(r) & \hbox{for black hole jet.} \\
                   \end{array} \right.
\end{equation}
where $\omega$ is the angular velocity associated with frame-dragging (see eqn.~\ref{eq:framedrag}) and 
\begin{equation}
\widetilde{\Omega}(r)={{\mathcal L}_{\rm ADAF}(r) \over r^2} {1\over  \gamma_r(r)\gamma_\phi(r)} \sqrt{{\mathcal{D}}(r) \over {\mathcal{A}}^{3}(r)},
\end{equation}
is angular frequency of the rotating fluid with respect to the local inertial observer, also often referred to as the zero angular momentum observer (ZAMO).   This is an observer who is dragged in azimuth by the spinning black hole and orbits with angular frequency $\omega$.  \scite{gampop98} refer to this observer's reference frame as the locally non-rotating frame.

To complete our calculation of the jet power, we need to estimate the magnitude of the large-scale poloidal component of the magnetic field, as seen by the observer at infinity in the Boyer-Lindquist coordinate system.   This field is related to that in the fluid rest frame by a boost factor $\gamma_\infty$:
\begin{equation}
B^{\rm LS}_{\rm pol,\infty} = \gamma_\infty B^{\rm LS}_{\rm pol, fluid}
\end{equation}
where \cite{takahashi02}:
\begin{equation}
\gamma_{\infty} = \gamma_r \gamma_\phi (2 j \beta_\phi/r^2+\sqrt{{\mathcal D}})/\sqrt{{\mathcal A}}.
\end{equation}

In our ADAF model, we assumed that the magnetic field is isotropically tangled.   The field in an accreting flow, however, is unlikely to be isotropic due to the presence of weak, ordered magnetic fields, and the results of MHD dynamo effects induced by turbulence as well as large-scale shearing motions.
\scite{livio99} (see also references therein) suggest that under such circumstances, the relationship between the toroidal and the poloidal components of the magnetic field can be approximated as  $B^{\rm LS}_{\rm pol, fluid} \approx {\mathcal{H}} B_{\rm \phi, fluid}$.   The resulting magnetic energy density is $U_{\rm mag}=(1+{\mathcal{H}}^2) B^2_{\rm \phi, fluid}/8\pi$.   Identifying this with the magnetic field energy density in our ADAF (i.e., $U_{\rm mag}=3(1-\beta)P_{\rm tot}$) and noting that ${\mathcal{H}} \approx 1/2$ at $R_{\rm ISCO}$  \cite{popgam98},  we can estimate $B^2_{\rm \phi, fluid}$.  The approach described here is based on the implicit assumption that the basic structure of the accretion flow will not change significantly by the implied presence of strongly ordered magnetic fields in the inner regions.   Recent numerical simulation results of \scite{beckwith08} suggest that this is a reasonable assumption.

Additionally, \scite{meier99} has argued that in the inner regions of the ADAF, the shear associated the differential dragging of reference frames will drive a dynamo that will draw on the poloidal component of the magnetic field  to generate and amplify the toroidal component  at the expense of the black hole's rotational energy.   This effect too has been tentatively observed in recent MHD numerical simulations \cite{hawley06}.
Denoting the corresponding field enhancement factor as  $g=\exp{(\omega\tau)}$,
\begin{equation}
(B^{\rm LS}_{\rm pol,\infty})^2 \approx \gamma^2 _\infty  { 24\pi \over 5} (1-\beta)P_{\rm tot} g^2.
\end{equation}

The amplitude of this enhancement factor depends on the black hole spin through the angular velocity of spacetime rotation, $\omega$.   In the case of a non-rotating black hole, $\omega=0$ and $g=1$ (i.e., no field enhancement).     The enhancement factor also depends on the characteristic timescale, $\tau$, over which the field amplification would occur.    The timescale typically associated with MHD processes is the orbital period of the rotating fluid as determined by the local inertial observer, $\tau = \tau_{\tilde{\phi}} \approx \widetilde{\Omega}^{-1}$, but doing so involves an implicit assumption that that radial motion of the fluid is negligible.   Such an approximation is reasonable if the characteristic radius of the jet is sufficiently far from the horizon; however, in the case of the black hole jet, the static limit is always close to the horizon. For the disk jet, as the black hole spin parameter tends towards unity, $R_{\rm ISCO}$ approaches the horizon, the radial velocity grows rapidly and the flow becomes increasingly radial in character. The rapid inflow will decrease the amount of time available for field enhancement. We take into account the resulting correction by modifying the characteristic timescale for field enhancement as follows:
\begin{equation}
\tau \approx  \tau_{\tilde{\phi}}\,\min\left[\,\, 1, \, {\tau_{\tilde{r}}\over \tau_{\tilde{\phi}}}\,\,\right]\approx \,\min\left[\,\, {1\over\widetilde{\Omega}}, \,\,{r \gamma_\phi \over \sqrt{\mathcal{D}}V}\,\,\right].
\label{eq:timescale}
\end{equation}
In the above equation, $\tau_{\tilde{r}}\approx (r/\sqrt{\mathcal{D}})/v^{\tilde{r}}$ is the characteristic inflow timescale in the local inertial frame.

\end{document}